\documentclass[12pt]{iopart}

\usepackage{iopams}
\usepackage{amsmath}
\usepackage{graphicx}
\usepackage{longtable}
\usepackage{cases}
\usepackage{mathrsfs}
\usepackage{dcolumn}
\usepackage{bm}
\usepackage{makecell}
\usepackage{epstopdf}
\begin{document}

\title{Extraction of the second and fourth radial moments of nuclear charge density from the elastic electron-nucleus scattering}
\author{Jian Liu$^{1,*}$, Xiaoting Liu$^{1,*}$, Xuezhi Wang$^{1}$, Shuo Wang$^{2}$, Chang Xu$^{3}$ and Zhongzhou Ren$^{4}$}
\address{$^{1}$College of Science,  China University of Petroleum (East China), Qingdao 266580, China}
\address{$^{2}$Shandong Provincial Key Laboratory of Optical Astronomy and Solar-Terrestrial Environment, Institute of Space Science, Shandong University, Weihai 264209, China}
\address{$^{3}$Department of Physics and Key Laboratory of Modern Acoustics, Nanjing University, Nanjing 210093, China}
\address{$^{4}$School of Physics Science and Engineering, Tongji University, Shanghai 200092, China}

\ead{*liujian@upc.edu.cn}
\ead{*Llxt0809@163.com}

\begin{abstract}
\par
\noindent

The expression for the radial moments $\left\langle r^{n}\right\rangle_{c}$ of the nuclear charge density has been discussed under the plane wave Born approximation (PWBA) method recently, which is significant to investigate the nuclear surface thickness and neutron distribution radius. In this paper, we extend the studies of extracting second-order moment $\left\langle r^{2}\right\rangle_{c}$ and fourth-order moment $\left\langle r^{4}\right\rangle_{c}$ from the Coulomb form factors $|F_{C}(q)|^2$ by the distorted wave Born approximation (DWBA) at the small momentum transfer $ q $ region. Based on the relativistic mean-field (RMF) calculations, the DWBA form factors $F_{C}^{DW}(q)$ are expanded into $ q^4 $, where the corresponding charge distributions are corrected by the contributions of neutron and spin-orbit densities. In the small $ q $ region, it is found that the experimental $|F_{C}(q)|^2$ can be well reproduced by considering the contributions of the $\left\langle r^{4}\right\rangle_{c}$ at the small $ q $ region. Through further analyzing the second-order and fourth-order expansion coefficients of the $F_{C}^{DW}(q)$, the relationship between the expansion coefficients and proton number $ Z $ is obtained. By the relationship, we extract the $\left\langle r^{2}\right\rangle_{c}$ and $\left\langle r^{4}\right\rangle_{c}$ from the limited experimental data of form factors at the small $ q $ region. Within the permissible range of error, the extracted $\left\langle r^{n}\right\rangle_{c}$ are consistent with the experimental data in this paper.

\end{abstract}

\submitto{\jpg}
\maketitle

\section{Introduction}\label{c1}
Electron scattering is an important method for the studies of the nuclear electromagnetic structure \cite{RevModPhys.28.214,PhysRevC.89.012201,de1966electron,RN11,walecka2001electron,uberall2012electron}, which is widely applied to extract precise nuclear information. It is noted that electromagnetic force is the main interaction between the electrons and the nuclei \cite{PhysRevC.95.044318,PhysRevC.96.034314}. Therefore, the nuclei are unperturbed during the elastic scattering process, and the data analysis is not affected by the uncertainties associated with the strong interaction \cite{PhysRevC.87.014304}. In the past few decades, plenty of elastic electron scattering experiments have been performed on stable nuclei \cite{ANGELI201369,DEVRIES1987495}. Supposing the nuclei are spherically symmetric, the nuclear charge density distributions and the charge radii are accurately extracted from the electron scattering data \cite{DEVRIES1987495}.

Electron scattering cross sections of nuclei are closely related to the radial moments $\left\langle r^{n}\right\rangle_{c}$ of the charge density distributions \cite{PhysRevA.94.042502}, which are defined as: 
\begin{align}\label{R}
	r_{n} \equiv \sqrt[n]{\left\langle r^{n}\right\rangle_{c}}=\left(\frac{\int d^{3} r\, r^{n} \rho_{c}(\mathbf{r})}{\int d^{3} r 
		\rho_{c}(\mathbf{r})}\right)^{1 / n}.
\end{align}
In general, the second radial moment $\left\langle r^{2}\right\rangle_{c}$ and the fourth radial moment $\left\langle r^{4}\right\rangle_{c}$ are more concerned. The $\left\langle r^{2}\right\rangle_{c}$ and $\left\langle r^{4}\right\rangle_{c}$ of charge density can reflect the nuclear electromagnetic properties to a certain extent. The charge root mean square (rms) radius can be obtained from the $\left\langle r^{2}\right\rangle_{c}$, which affects the locations of the minima of the cross sections \cite{RN156,RN154,RN171,PhysRevC.101.021301}. Besides, the $\left\langle r^{4}\right\rangle_{c}$ is another important nuclear property, which plays a significant role in investigating the nuclear structure. As a depiction of density diffusion around the nuclear surface, the $\left\langle r^{4}\right\rangle_{c}$ is strongly associated with the surface thickness $\sigma$ of the nuclear density distributions, and it also determines the diffraction radius $R$ of the heavy nuclei \cite{PhysRevC.101.021301,PhysRevC.79.034310,RN167,RN170,PhysRevC.33.335}. In addition to the surface thickness $\sigma$, the $\left\langle r^{4}\right\rangle_{c}$ noticeably influences the mean square radius $ R_n $ of the point neutron distributions, and the $ R_n^2 $ can be estimated based on the linear relationship between $\left\langle r^{4}\right\rangle_{c}$ and $ R_n^2 $ for the different nuclear models \cite{RN165}. In the past several decades, the $\left\langle r^{2}\right\rangle_{c}$ of stable nuclei have been precisely extracted from the Coulomb form factors $|F_{C}(q)|^2$ from the electron scattering experiments \cite{ANGELI201369,DEVRIES1987495}, and have been investigated extensively based on numerous nuclear structure models \cite{RN166,PhysRevC.72.044307,RN168,KARATAGLIDIS2007148,RN173}.

Recently, the elastic electron scattering experiment of $ ^{132} $Xe has been carried out firstly in the facility of the self-confining radioactive-isotope ion target (SCRIT) of RIKEN \cite{PhysRevLett.118.262501,PhysRevLett.100.164801,PhysRevLett.102.102501}. However, there is a limit to the incident energy of the experiment, the Coulomb form factors $|F_{C}(q)|^2$ were only measured in the region of $ q <1.5 \;\mathrm{fm}^{-1}$ \cite{PhysRevLett.118.262501}. For the purpose of connecting the experimental results with diverse orders of radial moments, in Ref. \cite{RN31}, the authors have systematically studied the $\left\langle r^{2}\right\rangle_{c}$ and $\left\langle r^{4}\right\rangle_{c}$ of the charge distributions by the plane wave Born approximation (PWBA) method. The results show that in the region of the small momentum transfer $q$, the $\left\langle r^{2}\right\rangle_{c}$ and $\left\langle r^{4}\right\rangle_{c}$ play the main role in the form factor \cite{RN31}. With the increase of $n$th-order moment, the influence of the $\left\langle r^{n}\right\rangle_{c}$ on the $|F_C(q)|^2$ gradually reduces at the small $ q $ region.

The PWBA method is a convenient approach to interpret the experimental data \cite{de1966electron,RN11}, where the $|F_C(q)|^2$ are expressed as the Fourier transformation of the density distributions. However, because the Coulomb distortion effects are neglected, the PWBA method calculations are not accurate enough, especially for the heavy nuclei \cite{RN112,PhysRevC.71.054323}. In order to calculate the $|F_C(q)|^2$ more adequately, the eikonal approximation \cite{PhysRev.134.B240} and phase shift analysis method \cite{NISHIMURA1985523,RN111,RN121} termed as the distorted wave Born approximation (DWBA) \cite{RN112} method are introduced by including the Coulomb distortion effects. Compared with the PWBA method, the $|F_C(q)|^2$ calculated by the DWBA method coincides better with the experimental data \cite{PhysRevC.72.044307,RN112}. In Refs. \cite{PhysRevC.73.014610,LIU201646,PhysRevC.89.014304,PhysRevC.79.014604,RN206,PhysRevC.82.024320,RN205,PhysRevC.98.044310,PhysRevC.70.034303,RN104}, the $|F_C(q)|^2$ are investigated systematically with the diverse nuclear structure model and the DWBA method.

This paper aims to extract the $\left\langle r^{2}\right\rangle_{c}$ and $\left\langle r^{4}\right\rangle_{c}$ of nuclear charge density distributions from the experimental $|F_C(q)|^2$ by the DWBA method at the small $q$ region. In the PWBA framework, it is proved that the $F_C(q)$ can be expanded to the different orders of $q^2$ at low momentum transfer region \cite{RN31}, and the coefficients of $ q^n $ are the $n$th-order moment of charge densities. For the DWBA method framework, we assume that the expansion coefficients of DWBA form factors $F_C^{DW}(q)$ can still be related to the radial moments. In this paper, we first calculate the $\left\langle r^{2}\right\rangle_{c}$ and $\left\langle r^{4}\right\rangle_{c}$ of nuclear charge density distributions from the relativistic mean-field (RMF) model with the $\mathrm N \mathrm L3 ^{*}$ and FSUGold parameter sets. The charge distributions also consider the contributions of the neutron density and the spin-orbit density. The sensitivities of the corrections for the neutron and spin-orbit densities on radial moments to the different parameterizations is discussed. Secondly, based on the charge densities, we calculate the $|F_C(q)|^2$ by the PWBA and DWBA methods, respectively. The comparisons between these two methods demonstrate the necessity of the DWBA method. Furthermore, we prove that at the small $q$ region, the expansion of DWBA form factors $F_C^{DW}(q)$ to the fourth-order are sufficient to describe the theoretical results. Finally, from light nuclei to medium mass nuclei, we calculate the expansion coefficients of the $F_C^{DW}(q)$ at the small $q$ region, and analyze the relationship between the expansion coefficients and proton number $ Z $. Based on the relationship, we extract the $\left\langle r^{2}\right\rangle_{c}$ and $\left\langle r^{4}\right\rangle_{c}$ from the experimental data. We extend the methodology from the PWBA method to the DWBA method for the extraction of $\left\langle r^{4}\right\rangle_{c}$ based on the limited experimental $|F_C(q)|^2$ at the small $q$ region. The study proposed in this paper can be used to interpret the subsequent experiments of elastic electron scattering on unstable nuclei, and can also offer useful guides for the coming experiments.

The paper is organized as follows: In Sec. \ref{c2}, the method for describing the relativistic nuclear charge density and the theoretical framework for obtaining the Coulomb form factors $|F_C(q)|^2$ with the PWBA and the DWBA methods are introduced. In Sec. \ref{c3}, the numerical results and discussions are presented. Finally, a summary is given in Sec. \ref{c4}.

\par
\section{Theoretical framework}\label{c2}

In this section, the nuclear charge density distributions are investigated from the relativistic mean-field (RMF) model, where both the contributions of nucleon density and spin-orbit density are taken into account. With the charge density distributions, the Coulomb form factors $|F_C(q)|^2$ are calculated by the PWBA and the DWBA methods, respectively. We further analyze the contributions of the $ n $th-order moment of nuclear charge densities to $|F_C(q)|^2$ at the small $ q $ region by the series expansion.

\subsection{Relativistic nuclear charge density}

In the previous studies, the charge distributions are mainly obtained by considering the spatial dispersion of the proton distributions \cite{SHARMA1993377,PhysRevC.55.540,CHABANAT1998231}. In this paper, we further take into account the contributions of the neutron density and the spin-orbit density to the nuclear charge density distributions. 

The nuclear charge density distributions from the RMF model are given by \cite{RN31}:
\begin{align}\label{rho c}
	\rho_{c}(r)=\sum_{\tau}\left(\rho_{c \tau}(r)+W_{c \tau}(r)\right),
\end{align}
where the nucleon charge density $\rho_{c\tau}$ and the spin-orbit charge density $W_{c\tau}$ can be obtained by:
\begin{subequations}
	\begin{align}\label{rho ct}
		\rho_{c \tau}(r)=\frac{1}{r} \int_{0}^{\infty} d x\, x \,\rho_{\tau}(x)\left(g_{\tau}(|r-x|)-g_{\tau}(r+x)\right), 	
	\end{align}
	\begin{align}\label{w ct}
		W_{c \tau}(r)=\frac{1}{r} \int_{0}^{\infty} d x\, x\, W_{\tau}(x)\left(f_{2 \tau}(|r-x|)-f_{2 \tau}(r+x)\right).
	\end{align}
\end{subequations}
The nucleon density $\rho_{\tau}$ and the spin-orbit density $W_{\tau}$ in the RMF model can be expressed as \cite{RN137}
\begin{subequations}
	\begin{align}  \label{rho t}
		\rho_{\tau}(r)=\sum_{\alpha \in \tau} \frac{2 j_{\alpha}+1}{4 \pi r^{2}}\left(G_{\alpha}^{2}+F_{\alpha}^{2}\right),
	\end{align}	
	\begin{align}\label{w t}
		W_{\tau}(r)=\frac{\mu_{\tau}}{M} \sum_{\alpha \in \tau} \frac{2 j_{\alpha}+1}{4 \pi r^{2}} \frac{d}{d r}\left(\frac{M-M^{*}(r)}{M} G_{\alpha} F_{\alpha}+\frac{\kappa_{\alpha}+1}{2 M r} G_{\alpha}^{2}-\frac{\kappa_{\alpha}-1}{2 M r} F_{\alpha}^{2}\right).
	\end{align}
\end{subequations}
The $\mu_\tau$ is the anomalous magnetic moment. The Eq. (\ref{rho t}) satisfies the normalization condition $\int d^{3} r \rho_{\tau}(r)=Z, N$ for $\tau = p, n$, respectively, while the Eq. (\ref{w t}) satisfies $\int d^{3} r W_{\tau}(r)=0$. The functions $g_{\tau}(x)$ and $f_{2 \tau}(x)$ are 
	\begin{align}\label{gt f2t}
		g_{\tau}(x)=\frac{1}{2 \pi} \int_{-\infty}^{\infty} d q\, e^{i q x} G_{E \tau}\left(q^{2}\right),\  \
		f_{2 \tau}(x)=\frac{1}{2 \pi} \int_{-\infty}^{\infty} d q\, e^{i q x} F_{2 \tau}\left(q^{2}\right),
	\end{align}
where the $G_{E \tau}\left(q^{2}\right)$ and $F_{2 \tau}\left(q^{2}\right)$ are the Sachs and Pauli form factors, respectively . The $G_{E \tau}\left(q^{2}\right)$ and $F_{2 \tau}\left(q^{2}\right)$ have different prescriptions \cite{RN31,RN137,BERTOZZI1972408,RN143}, and in our calculations the specific expressions on $G_{E \tau}\left(q^{2}\right)$ and $F_{2 \tau}\left(q^{2}\right)$ of Ref. \cite{RN31} are taken into account.

In the previous study \cite{BERTOZZI1972408}, it has been shown that neutrons have important contributions to elastic electron scattering, because the point neutron also has a spatial charge distribution. It has been explained in Ref. \cite{RN31} that there are impressive influences of neutrons on the charge distributions. Therefore, it is necessary and significant to take into account the correction of neutrons during the calculations of charge distributions. 

The neutron spin-orbit density contributes to the nuclear charge distribution as a relativistic effect. The contribution is enhanced by the effective mass stemming from the Lorentz-scalar potential in relativistic. The spin-orbit contributions in nuclear charge density are not adequately described in previous mean-field calculations and some meaningful work discusses those impacts \cite{RN137,RN203,RN204,RN195}. Therefore, in our calculations we take into account the corrections of the spin-orbit density in Eq. (\ref{rho c}). 

\subsection{The $F_C^{PW}(q)$ from the PWBA method}

Based on Eq. (\ref{rho c}), the Coulomb form factor can be calculated by the PWBA method. In the PWBA method, the form factor is calculated from the nuclear charge density distributions via the Fourier transform:
\begin{align}\label{fc PW}
	F_{C}^{P W}(q)=\frac{1}{Z} \int \rho_{c}(\mathbf{r}) \,e^{i \mathbf{q} \cdot \mathbf{r}} d \mathbf{r}.
\end{align}
When calculating the PWBA form factors, the Coulomb distortion can be taken into account approximately by replacing the momentum transfer $ q $ with the effective momentum transfer $ q_\mathrm{eff} $ \cite{PhysRevC.72.044307,NISHIMURA1985523,RN194}:
\begin{align}\label{qeff}
    q_\mathrm{eff}=q\left[1+1.5 \alpha Z \hbar c /\left(E R_{0}\right)\right],
\end{align}
where $ E $ is the incident energy, $ R_0=1.07A^{1/3} $ and $ A $ is the nuclear mass number. The exponential function can be expanded in series:
\begin{align}\label{eiqr}
	e^{i \mathbf{k} \cdot \mathbf{r}}=\sum_{l=0}^{\infty} i^{l}(2 l+1) j_{l}(k r) P_{l}(\cos \theta).
\end{align}
Substituting Eq. (\ref{eiqr}) into the Eq. (\ref{fc PW}) and considering the spherical symmetry, we can obtain the contributions of the $n$th-order moment of the $\rho_{c}(r)$ to the PWBA form factor $F_{C}^{P W}(q)$:
\begin{align}\label{Fc PW}
	F_{C}^{P W}(q)=1-\frac{1}{6} q_\mathrm{eff}^{2}\left\langle r^{2}\right\rangle+\frac{1}{120} q_\mathrm{eff}^{4}\left\langle r^{4}\right\rangle+\ldots+(-1)^{n} \frac{q_\mathrm{eff}^{2 n}\left\langle r^{2 n}\right\rangle}{(2 n+1) !},
\end{align}
where the $n$th-order moment is defined in Eq. (\ref{R}).

\subsection{The $F_C^{DW}(q)$ from the DWBA method}

The PWBA method is a convenient approach for the light nuclei, but it is less accurate for the medium nuclei and heavy nuclei. Due to the enhancement of the nuclear electromagnetic field, the wave functions of the scattered electrons are distorted. The DWBA method is more accurate in the case of heavy nuclei because the Coulomb distortion effects are further taken into account.

In the DWBA method, the wave functions of the scattered electrons can be calculated by the Dirac equation \cite{RN145,RN80,bjorken1964relativistic}:
\begin{align}\label{Dirac}
	[\alpha \cdot \mathbf{p}+\beta m+V(\mathbf{r})] \Psi(\mathbf{r})=E \Psi(\mathbf{r}).
\end{align}
The potential energy of an electron at a distance $ r $ from the center of the nucleus is given by \cite{PhysRevC.78.044332}:
\begin{align}\label{V r}
    V(r)=-4 \pi e\left[\frac{1}{r} \int_{0}^{r} \rho_{ch}\left(r^{\prime}\right) r^{\prime 2} d r^{\prime}+\int_{r}^{\infty} \rho_{ch}\left(r^{\prime}\right) r^{\prime} d r^{\prime}\right],
\end{align}
where $ \rho_{ch}(r) $ denotes the charge density of the nucleus, considered to be spherically symmetrical.

By resolving the Dirac equation with the scattering boundary condition \cite{RN146}, the phase shifts of spin-up ones $\delta_l^+$ and spin-down ones $\delta_l^-$ can be obtained. The direct and the spin-flip scattering amplitude can be determined by \cite{PhysRevC.82.024320,PhysRevC.78.044332,PhysRevC.79.044313}:
\begin{subequations}
	\begin{align}\label{f sita}
		f(\theta)=\frac{1}{2 i k} \sum_{l=0}^{\infty}\left[(l+1)\left(e^{2 i \delta_{t}^{+}}-1\right)+l\left(e^{2 i \delta_{l}^{-}}-1\right)\right] P_{l}(\cos \theta),
	\end{align}
	\begin{align}\label{g sita}
		g(\theta)=\frac{1}{2 i k} \sum_{l=0}^{\infty}\left[e^{2 i \delta_{l}^{-}}-e^{2 i \delta_{l}^{+}}\right] P_{l}^{1}(\cos \theta),
	\end{align}
\end{subequations}
where the $P_l$ and $P_l^1$ represent the Legendre functions and associated Legendre functions, respectively.

The scattering cross sections are denoted as
\begin{align}\label{cross sections}
	\frac{d \sigma}{d \Omega}=|f(\theta)|^{2}+|g(\theta)|^{2},
\end{align}
as well as the form factors:
\begin{align}\label{DW ffs}
	\left|F_{C}{ }^{D W}(q)\right|^{2}=\frac{d \sigma / d \Omega}{d \sigma_{M} / d \Omega}.
\end{align}
The Mott cross section $ \frac{d \sigma_{M}}{d \Omega}=\left(\frac{Z \alpha^{2}}{2 E}\right)^{2} \frac{\cos ^{2} \frac{\theta}{2}}{\sin ^{4} \frac{\theta}{2}} $. With the Eqs. (\ref{Dirac}) - (\ref{DW ffs}), one can finally obtain the DWBA form factors.

According to Eq. (\ref{Fc PW}), we expand the $F_C^{DW}(q)$ obtained by the DWBA method into series of $q^2$ at the small $ q $ region:
\begin{align}\label{Fc DW}
	F_{C}^{D W}(q)=\sum_{n} D_{2 n} \frac{(-1)^{n} q^{2 n}\left\langle r^{2 n}\right\rangle}{(2 n+1) !},
\end{align}
where
\begin{align}\label{D2n}
	D_{2 n}=\left.\frac{\partial^{2 n} F_C^{DW}(q)}{\partial q^{2 n}}\right|_{q=0} \frac{(-1)^{n}(2 n+1)}{\left\langle r^{2 n}\right\rangle}.
\end{align}
The expansion coefficients $D_{2n}$ illustrate the contributions of the $n$th-order moment of the $\rho_{c}(r)$ in the DWBA method. When directly calculating the derivative of Eq. (\ref{D2n}), the error of the calculated $ D_{2n} $ is a little larger. Therefore, the polynomial fitting is applied to obtain the $ D_{2n} $:
\begin{align}\label{Fc DW nihe}
    F_{C}^{D W}(q)=\sum_{n=0,1,2} a_{2 n} q^{2 n}.
\end{align}
With the fit coefficients $ a^{2n} $, the expansion coefficients $ D_{2n} $ in Eq. (\ref{D2n}) can be rewritten as
\begin{align}\label{D2n nihe}
	D_{2 n}=a_{2 n} \frac{(-1)^{n}(2 n+1) !}{\left\langle r^{2 n}\right\rangle}.
\end{align}
The expansion formulas Eqs. (\ref{Fc PW}) and (\ref{Fc DW}) for the PWBA and DWBA form factors are only valid at the small $ q $ region, which is before the first minimum of the form factor. For electron scattering experiments on different nuclei, the measurements of the small $ q $ region are mainly located at the range $0.3 \;\mathrm{ fm}^{-1}<q<0.7 \;\mathrm{fm}^{-1}$. Therefore, the fourth-order polynomial fitting is taken for the DWBA form factors in the region $0.3 \;\mathrm{fm}^{-1}<q<0.7 \;\mathrm{fm}^{-1}$ to obtain the expansion coefficients $D_{2n}$.

\section{Numerical results and discussion}\label{c3}
In this section, the second-order moment $\left\langle r^{2}\right\rangle_{c}$ and fourth-order moment $\left\langle r^{4}\right\rangle_{c}$ of the charge density distributions and the Coulomb form factors $|F_{C}(q)|^2$ are investigated with the formulas of Sec. \ref{c2}. The corresponding nuclear charge density distributions are obtained from the RMF model with the $\mathrm N \mathrm L3 ^{*}$ \cite{RN196} and $\mathrm{FSUGold}$ \cite{RN197} parameter sets ($\mathrm{FSU}$ for short in the following sections). Furthermore, we discuss the relationship between DWBA form factors $|F_{C}^{DW}(q)|^2$ and radial moments at the small momentum transfer $q$ region. On the basis of this relationship, the $\left\langle r^{2}\right\rangle_{c}$ and $\left\langle r^{4}\right\rangle_{c}$ of the nuclei charge densities are extracted with the limited experimental data at the small $ q $ region.

\subsection{Contributions of neutron and spin-orbit densities to the $\rho_{c}(r)$}

Combining Eqs. (\ref{R}) and (\ref{rho c}), one can get the expression for the $n$th-order moment of charge distributions $\left\langle r^{2 n}\right\rangle_{c}$ as follows:
\begin{align}\label{r2nc}
	\left\langle r^{2 n}\right\rangle_{c}=\left\langle r^{2 n}\right\rangle_{\rho_{c p}}+\left\langle r^{2 n}\right\rangle_{\rho_{c n}}+\left\langle r^{2 n}\right\rangle_{W_{c p}}+\left\langle r^{2 n}\right\rangle_{W_{c n}},
\end{align}
where
   \begin{align}\label{r2npct}
		\left\langle r^{2 n}\right\rangle_{\rho_{c\tau}}=\frac{\int \rho_{c \tau}(r) r^{2 n} d^{3} r}{\int \rho_{c}(r) d^{3} r}, \ \
		\mathrm{and} \quad
		\left\langle r^{2 n}\right\rangle_{W_{c\tau}}=\frac{\int W_{c \tau}(r) r^{2 n} d^{3} r}{\int \rho_{c}(r) d^{3} r}.
   \end{align}
The $ \left\langle r^{2 n}\right\rangle_{\rho_{c\tau}} $  present the contributions from the space dispersion of the point proton and neutron densities, and the $ \left\langle r^{2 n}\right\rangle_{W_{c\tau}} $ are the contributions of the spin-orbit densities of proton and neutron, respectively.


In table \ref{tableI}, we present the theoretical $ \left\langle r^{2}\right\rangle_{\rho_{c p}} $, $ \left\langle r^{2}\right\rangle_{\rho_{c n}} $, $ \left\langle r^{2}\right\rangle_{W_{c p}} $, $ \left\langle r^{2}\right\rangle_{W_{c n}} $, and $ \left\langle r^{2}\right\rangle_{c} $ of the charge density distributions calculated from RMF model with $\mathrm{NL3}^{*}$ and FSU parameter sets, respectively. For $^{40,48}$Ca, the experimental $ \left\langle r^{2}\right\rangle_{c} $ are calculated by the Fourier-Bessel (FB) density \cite{DEVRIES1987495}, and for $^{116,124}$Sn, those are calculated from the Sum-of-Gaussians (SOG) density \cite{DEVRIES1987495}.

\begin{table}[b]
	\caption{The second-order moment $\left\langle r^{2}\right\rangle_{c}$ of the charge density distributions of $^{40,48}$Ca and $^{116,124}$Sn from the RMF model with the $\mathrm{NL3}^{*}$ and $\mathrm{FSU}$ parameter sets. The quantities are calculated with Eq. (\ref{r2nc}) in units of $\mathrm{fm}^{2}$. The experimental data are from Ref. \cite{DEVRIES1987495}. }
	\renewcommand{\arraystretch}{1.1}
	\centering
	\begin{tabular}{p{1.3cm}p{1.6cm}p{1.6cm}p{1.6cm}p{1.6cm}p{1.6cm}p{1.6cm}p{1.3cm}}
		\hline
		\hline
		Nuclei & Model   & $\left\langle r^{2}\right\rangle_{\rho_{c p}}$ & $\left\langle r^{2}\right\rangle_{\rho_{c n}}$ & $\left\langle r^{2}\right\rangle_{W_{c p}}$ & $\left\langle r^{2}\right\rangle_{W_{c n}}$ & $\left\langle r^{2}\right\rangle_{c}$ & Expt.  \\ \hline
		$^{40}$Ca   & $\mathrm{NL3}^{*}$ & 12.059                         & -0.1200                        & 0.0220                         & -0.0242                        & 11.937                       & 11.902 \\
		& $\mathrm{FSU}$ & 11.798                         & -0.1200                        & 0.0243                         & -0.0268                        & 11.676                       &        \\
		$^{48}$Ca   & $\mathrm{NL3}^{*}$ & 12.061                         & -0.1680                        & 0.0260                         & -0.1565                        & 11.763                       & 11.910 \\
		& $\mathrm{FSU}$ & 11.985                         & -0.1680                        & 0.0279                         & -0.1626                        & 11.682                       &        \\
		$^{116}$Sn  & $\mathrm{NL3}^{*}$ & 21.201                         & -0.1584                        & 0.1091                         & -0.0566                        & 21.095                       & 21.405 \\
		& $\mathrm{FSU}$ & 21.227                         & -0.1584                        & 0.1118                         & -0.0600                        & 21.120                       &        \\
		$^{124}$Sn  & $\mathrm{NL3}^{*}$ & 21.693                         & -0.1776                        & 0.1112                         & -0.1601                        & 21.467                       & 21.873 \\
		& $\mathrm{FSU}$ & 21.840                         & -0.1750                        & 0.1135                         & -0.1663                        & 21.612                       &        \\ \hline
		\hline
	\end{tabular}
	\label{tableI}
\end{table}

Table \ref{tableI} presents the contributions of each term to the $ \left\langle r^{2}\right\rangle_{c} $ for light isotopes $^{40,48}$Ca and heavy isotopes $^{116,124}$Sn. It can be seen from table \ref{tableI} that the $ \left\langle r^{2}\right\rangle_{\rho_{c n}} $, $ \left\langle r^{2}\right\rangle_{W_{c p}} $, $ \left\langle r^{2}\right\rangle_{W_{c n}} $ have significant corrections to the first term $ \left\langle r^{2}\right\rangle_{\rho_{c p}} $. For $^{40,48}$Ca, the proton quadratic moment $ \left\langle r^{2}\right\rangle_{\rho_{c p}} $ increases with the increasing of neutron number. By considering the corrections from neutron and spin-orbit densities, the $ \left\langle r^{2}\right\rangle_{c} $ decrease as the neutron number increases. The reason is that the neutron quadratic moment $ \left\langle r^{2}\right\rangle_{\rho_{c n}} $ brings in a negative contribution to the $ \left\langle r^{2}\right\rangle_{c} $. For the nucleus with proton number $ Z $ equal to neutron number $ N $, the contributions from the proton spin-orbit density $ \left\langle r^{2}\right\rangle_{W_{c p}} $ and neutron spin-orbit density $ \left\langle r^{2}\right\rangle_{W_{c n}} $ cancel each other out. For neutron-rich nuclei, the contributions of $ \left\langle r^{2}\right\rangle_{W_{c n}} $ to $ \left\langle r^{2}\right\rangle_{c} $ are important. Therefore, the corrections of the $ \left\langle r^{2}\right\rangle_{c} $ by the spin-orbit density and neutron density of the neutron-rich nuclei cannot be ignored. 

In order to discuss the sensitivities of the corrections to the parameterizations, the theoretical $ \left\langle r^{2}\right\rangle_{c} $ calculated from $\mathrm{NL3}^{*}$ and $\mathrm{FSU}$ parameter sets are also compared with each other in table \ref{tableI}. One can see that there are noticeable discrepancies on the $ \left\langle r^{2}\right\rangle_{\rho_{c p}} $ for different parameter sets, which is the main contribution to $ \left\langle r^{2}\right\rangle_{c} $. For certain nuclei, the values of $ \left\langle r^{2}\right\rangle_{\rho_{c n}} $, $ \left\langle r^{2}\right\rangle_{W_{c p}} $, $ \left\langle r^{2}\right\rangle_{W_{c n}} $ calculated from two parameter sets are close. This is because of the subtraction of $g_n(|r-x|)- g_n(r+x)$ and the subtraction of $f_{2\tau}(|r-x|)- f_{2\tau}(r+x)$ in Eqs. (\ref{rho ct}) and (\ref{w ct}) are also small and almost similar for the different parameters. Therefore, the corrections from the $ \left\langle r^{2}\right\rangle_{\rho_{c n}} $, $ \left\langle r^{2}\right\rangle_{W_{c p}} $, $ \left\langle r^{2}\right\rangle_{W_{c n}} $ are insensitive to different parameterizations.

We also calculate the $ \left\langle r^{4}\right\rangle_{\rho_{c p}} $, $ \left\langle r^{4}\right\rangle_{\rho_{c n}} $, $ \left\langle r^{4}\right\rangle_{W_{c p}} $, $ \left\langle r^{4}\right\rangle_{W_{c n}} $, and $ \left\langle r^{4}\right\rangle_{c} $ of the charge density distributions with $\mathrm{NL3}^{*}$ and FSU parameter sets, and the results are shown in table \ref{tableII}. For comparison, the corresponding experimental data \cite{DEVRIES1987495} are also presented.

\begin{table}[t]
	\caption{The fourth-order moment $\left\langle r^{4}\right\rangle_{c}$ of the charge density distributions of $^{40,48}$Ca and $^{116,124}$Sn predicted by the RMF model with the $\mathrm{NL3}^{*}$ and $\mathrm{FSU}$ parameter sets. The value of each term is calculated with Eq. (\ref{r2nc}) in units of $\mathrm{fm}^{4}$. The experimental data are from Ref. \cite{DEVRIES1987495}. }
	\renewcommand{\arraystretch}{1.1}
	\centering
	\begin{tabular}{p{1.3cm}p{1.6cm}p{1.6cm}p{1.6cm}p{1.6cm}p{1.6cm}p{1.6cm}p{1.3cm}}
		\hline
		\hline
		Nuclei & Model   & $\left\langle r^{4}\right\rangle_{\rho_{c p}}$ & $\left\langle r^{4}\right\rangle_{\rho_{c n}}$ & $\left\langle r^{4}\right\rangle_{W_{c p}}$ & $\left\langle r^{4}\right\rangle_{W_{c n}}$ & $\left\langle r^{4}\right\rangle_{c}$ & Expt.   \\ \hline
		$^{40}$Ca                       & $\mathrm{NL3}^{*}$ & 210.209                        & -4.918                         & 0.493                          & -0.559                         & 205.225                      & 199.991 \\
		& $\mathrm{FSU}$ & 202.376                        & -4.807                         & 0.613                          & -0.688                         & 197.494                      &         \\
		$^{48}$Ca                       & $\mathrm{NL3}^{*}$ & 204.010                        & -7.971                         & 0.779                          & -5.255                         & 191.563                      & 194.714 \\
		& $\mathrm{FSU}$ & 201.677                        & -7.790                         & 0.884                          & -5.433                         & 189.338                      &         \\
		$^{116}$Sn                      & $\mathrm{NL3}^{*}$ & 588.857                        & -12.360                        & 5.941                          & -3.180                         & 579.258                      & 603.180 \\
		& $\mathrm{FSU}$ & 589.850                        & -12.132                        & 6.158                          & -3.550                         & 580.326                      &         \\
		$^{124}$Sn                      & $\mathrm{NL3}^{*}$ & 612.953                        & -14.748                        & 6.231                          & -9.939                         & 594.497                      & 624.196 \\
		& $\mathrm{FSU}$ & 618.528                        & -14.128                        & 6.459                          & -10.407                         & 600.452                      &         \\ \hline
		\hline
	\end{tabular}
    \label{tableII}
\end{table}

From table \ref{tableII}, one can see that the RMF model can provide reasonable descriptions for the $ \left\langle r^{4}\right\rangle_{c} $ of the charge density distributions. In table \ref{tableII}, the first term $ \left\langle r^{4}\right\rangle_{\rho_{c p}} $ provides the dominant contributions for the value of $ \left\langle r^{4}\right\rangle_{c} $. A negative contribution of the neutron density from the second term $ \left\langle r^{4}\right\rangle_{\rho_{c n}} $ can cancel out the partial value of proton quartic moment $ \left\langle r^{4}\right\rangle_{\rho_{c p}} $. The contributions of neutron spin-orbit quartic moment $ \left\langle r^{4}\right\rangle_{W_{c n}} $ to the neutron-rich nuclei are noticeable. Therefore, in the neutron-rich nuclei, the corrections of the neutron density and spin-orbit density to the $ \left\langle r^{4}\right\rangle_{c} $ need to be taken into account.

Similar to the $ \left\langle r^{2}\right\rangle_{c} $, we also analyze the sensitivities of corrections on $ \left\langle r^{4}\right\rangle_{c} $ for different parameter sets. Compared with the differences between $ \left\langle r^{4}\right\rangle_{\rho_{c p}} $ from two parameter sets, the discrepancies on other corrections from two parameter sets are small, which shows again that the proton density provides the dominant contribution to $ \left\langle r^{4}\right\rangle_{c} $. This is also due to the small and similar contributions of $g_n(|r-x|)- g_n(r+x)$ and $f_{2\tau}(|r-x|)- f_{2\tau}(r+x)$ to different parameters in Eqs. (\ref{rho ct}) and (\ref{w ct}). Therefore, the corrections from $ \left\langle r^{4}\right\rangle_{\rho_{c n}} $, $ \left\langle r^{4}\right\rangle_{W_{c p}} $, $ \left\langle r^{4}\right\rangle_{W_{c n}} $ on $ \left\langle r^{4}\right\rangle_{c} $ are also not sensitive to the different parameters.

Besides the corrections of neutron and spin-orbit, the center-of-mass (c.m.) correction also affects the nuclear radial moments by unfolding with the width of the center-of-mass vibrations \cite{RN198,RN199,RN200}. In Ref. \cite{RN191}, the contributions of each term to the radial moments are calculated and presented visually. One can see that the contribution of c.m. correction is very small compared with the contributions of protons, neutrons, and spin-orbit. The c.m. correction accounts for only one-fifth of the neutron contribution. Therefore, the c.m. correction to the radial moments is ignored in our calculations.

It should be mentioned that in tables \ref{tableI} and \ref{tableII}, the final results of $ \left\langle r^{2}\right\rangle_{c} $ and $ \left\langle r^{4}\right\rangle_{c} $ in the seventh column are closer to the experimental values than the results of $ \left\langle r^{2}\right\rangle_{\rho_{c p}} $ and $ \left\langle r^{4}\right\rangle_{\rho_{c p}} $ in the third column for $^{40,48}$Ca, while for $^{116,124}$Sn it is just the opposite. This can be attributed to the theoretical framework of the RMF model. Under the no-sea approximations and mean-field approximations, the calculated $ \left\langle r^{2}\right\rangle_{\rho_{c p}} $ and $ \left\langle r^{4}\right\rangle_{\rho_{c p}} $ of the RMF model are more accurate and reliable for heavy nuclei, compared with those for light nuclei. If the parameters of the RMF model are not adjusted and the contributions from neutron density and spin-orbit density are added to the charge density, this leads to improvements in the descriptions on the radial moments of light nuclei and the opposite trend for heavy nuclei.

\subsection{The expansion of $|F_{C}(q)|^2$ for the nuclei}\label{c3.2}

The second-order moment $ \left\langle r^{2}\right\rangle_{c} $ and fourth-order moment $ \left\langle r^{4}\right\rangle_{c} $ of charge density only roughly describe the electromagnetic properties of nuclei. Instead of the $ \left\langle r^{2}\right\rangle_{c} $ and $ \left\langle r^{4}\right\rangle_{c} $, the Coulomb form factors $|F_{C}(q)|^2$ can more exactly reflect the nuclear electromagnetic structures. Therefore, in this part the $|F_{C}(q)|^2$ are further studied by the PWBA and DWBA methods, and the theoretical results are compared with the experimental data, respectively. 

\begin{figure}[t]
	\centering
	\includegraphics[width=14.cm,angle=0]{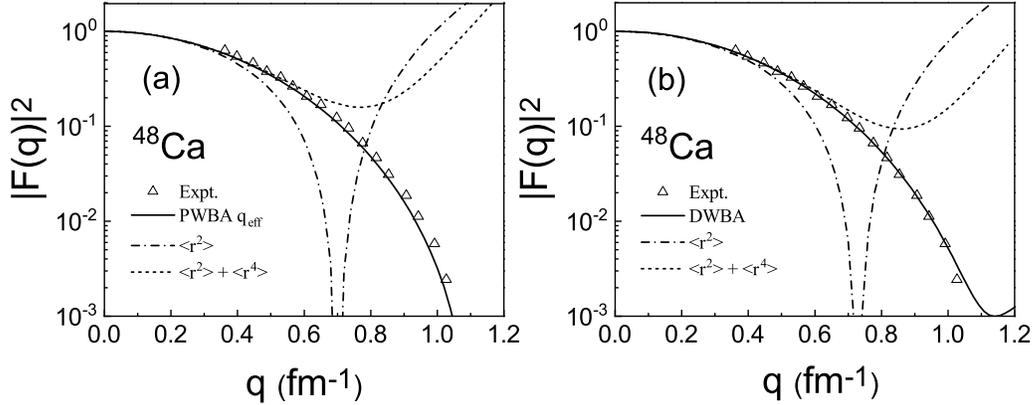}\\
	\caption{  (a) The contributions of the second-order moment $ \left\langle r^{2}\right\rangle_{c} $ and fourth-order moment $ \left\langle r^{4}\right\rangle_{c} $ of the charge density distributions to the form factors for $^{48}$Ca in the PWBA method from the RMF model with the $\mathrm{NL3}^{*}$ parameter set. The PWBA form factors are calculated via $ q_\mathrm{eff} $ of Eq. (\ref{qeff}). 
		\label{figure:48Ca a} 
		(b) The same as figure \ref{figure:48Ca a}(a), but for the DWBA method. The experimental data are taken from Ref. \cite{PhysRevC.67.034317}. 
		\label{figure:48Ca b}}
	\label{figure:48Ca}
\end{figure}

With the charge distributions of Eq. (\ref{rho c}) calculated by the RMF model with $\mathrm{NL3}^{*}$ parameter set, the corresponding $|F_{C}(q)|^2$ of $^{48}$Ca are investigated at the small momentum transfer $q$ region by the PWBA and the DWBA methods, and the results are presented in figures \ref{figure:48Ca a}(a) and \ref{figure:48Ca b}(b), respectively. In figure \ref{figure:48Ca a}(a), the solid line represents the $|F_{C}(q)|^2$ calculated from the PWBA method with  Eq. (\ref{fc PW}), where the $ q_\mathrm{eff} $ is used instead of $ q $ with  Eq. (\ref{qeff}). By this way, the minima of the form factor $ F_{C}^{PW}(q) $ can agree with the experimental ones better. One can see the theoretical results of $|F_{C}(q)|^2$ of $^{48}$Ca calculated by two methods are both consistent with the experimental data. This means that for light nuclei, both the PWBA and DWBA methods can reproduce the experimental $|F_{C}(q)|^2$ at the small $q$ region.

In figure \ref{figure:48Ca}(a), we expand the PWBA form factor $ F_C^{PW}(q) $ to $q^4$ by Eq. (\ref{Fc PW}), which presents the contributions of $\left\langle r^{n}\right\rangle_{c}$ of charge distributions. For the DWBA form factor $ F_C^{DW}(q) $ in figure \ref{figure:48Ca}(b), the fourth-order polynomial fitting is used in the small $ q $ region to obtain the contributions of $\left\langle r^{n}\right\rangle_{c}$ of charge distributions. By the Eqs. (\ref{Fc DW nihe}) and (\ref{D2n nihe}), we obtain $ D_2 =0.9472$ and $ D_4=0.6683$ for $ F_C^{DW}(q) $ of figure \ref{figure:48Ca}(b). Substituting the $ D_2 $ and $ D_4 $ into Eq. (\ref{Fc DW}), the contributions of $\left\langle r^{2}\right\rangle_{c}$ and $\left\langle r^{2}\right\rangle_{c}$+$\left\langle r^{4}\right\rangle_{c}$ for DWBA form factor in figure \ref{figure:48Ca}(b) can be obtained.

From figures \ref{figure:48Ca a}(a) and \ref{figure:48Ca b}(b), one can see that the contributions of $\left\langle r^{2}\right\rangle_{c}$ dominates the $|F_{C}(q)|^2$ of $^{48}$Ca up to $q \approx 0.3 \;\mathrm{fm}^{-1}$. In the region where $q > 0.3 \;\mathrm{fm}^{-1}$, there is a significant deviation between contributions of $ \left\langle r^{2}\right\rangle_{c} $ and $|F_{C}(q)|^2$. When further considering the contributions of $ \left\langle r^{4}\right\rangle_{c} $, the results agree with the experimental data up to $q \approx 0.6 \;\mathrm{fm}^{-1}$ in figures \ref{figure:48Ca a}(a) and \ref{figure:48Ca b}(b). Therefore, for light nuclei, there is little difference between PWBA and DWBA form factors at the small $ q $ region. When taking into account the contributions of $ \left\langle r^{4}\right\rangle_{c} $ of the charge density, the form factors from the PWBA method in Eq. (\ref{Fc PW}) and DWBA method in Eq. (\ref{Fc DW}) can all reproduce the experimental data in the $q$ region $0\sim 0.6 \;\mathrm{fm}^{-1}$. 

\begin{figure}[t]
	\centering
	\includegraphics[width=14.cm,angle=0]{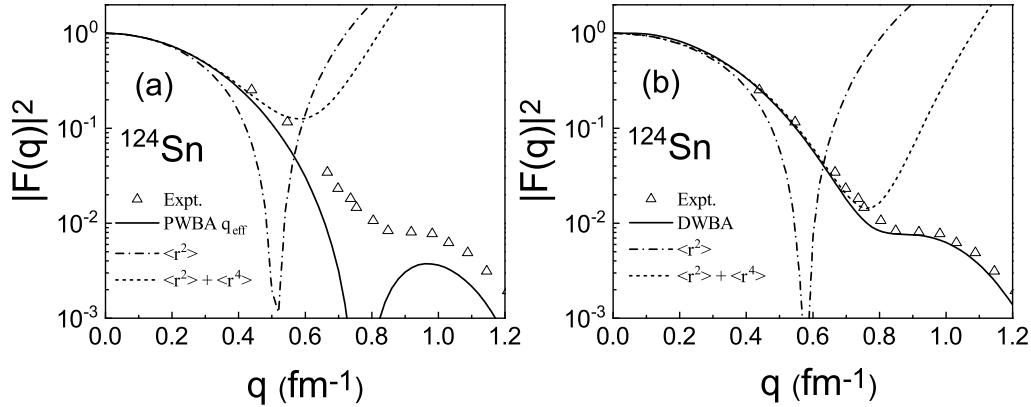}\\
	\caption{  (a) The contributions of the second-order moment $ \left\langle r^{2}\right\rangle_{c} $ and fourth-order moment $ \left\langle r^{4}\right\rangle_{c} $ of the charge density distributions to the form factors for $^{124}$Sn in the PWBA method from the RMF model with the $\mathrm{NL3}^{*}$ parameter set. The PWBA form factors are calculated via $ q_\mathrm{eff} $ of Eq. (\ref{qeff}).
		\label{figure:124Sn a} 
		(b) The same as figure \ref{figure:124Sn a}(a), but for the DWBA method. The experimental data are taken from Ref. \cite{RN148}. 
		\label{figure:124Sn b}}
	\label{figure:124Sn}
\end{figure}

Besides $^{48}$Ca, the $|F_{C}(q)|^2$ of $^{124}$Sn are also investigated by the PWBA and DWBA methods at the small $q$ region where the corresponding charge distributions are calculated by the RMF model with $\mathrm{NL3}^{*}$ parameter set. The results are presented in figures \ref{figure:124Sn a}(a) and \ref{figure:124Sn b}(b), respectively. In figure \ref{figure:124Sn a}(a), the $ q_\mathrm{eff} $ is also used when computing the PWBA form factors, which make the position of minima of the PWBA form factors coincide with the experimental data better. In this figure, there are some differences between the PWBA and DWBA form factors for $ ^{124} $Sn. One can see the values of $ |F_C^{PW}(q)|^2 $ deviate from the experimental data.  The DWBA method corrects this problem, and the results coincide with the experimental data well. This is due to the nuclear electromagnetic field of heavy nuclei, which distorts the wave function of scattered electrons. Therefore, it is more accurate to use the DWBA method to calculate the $|F_{C}(q)|^2$ of heavy nuclei.

In figure \ref{figure:124Sn a}(a), the PWBA form factors are also expanded to $ q^4 $ by Eq. (\ref{Fc PW}) at the small $ q $ region, which displays the contributions of $\left\langle r^{2}\right\rangle_{c}$ and $\left\langle r^{4}\right\rangle_{c}$ for heavy nuclei. In figure \ref{figure:124Sn b}(b), we fit the DWBA form factors $ F_C^{DW}(q) $ of small $ q $ region by the fourth-order polynomials Eq. (\ref{Fc DW nihe}) and obtain the coefficients $ D_2=0.8294 $ and $ D_4=0.4878 $ for $ ^{124} $Sn by Eq. (\ref{D2n nihe}). Substituting the $ D_2 $ and $ D_4 $ into Eq. (\ref{Fc DW}), the contributions of $\left\langle r^{2}\right\rangle_{c}$ and $\left\langle r^{2}\right\rangle_{c}$+$\left\langle r^{4}\right\rangle_{c}$ for DWBA form factor of $ ^{124} $Sn in figure \ref{figure:124Sn b}(b) can be provided.

From figure \ref{figure:124Sn a}(a), it can be seen that for the PWBA method, taking into account the contributions of $\left\langle r^{2}\right\rangle_{c}$ and $\left\langle r^{4}\right\rangle_{c}$ can only describe the form factors in the region of $q \approx 0\sim0.4 \;\mathrm{fm}^{-1}$. For the DWBA method in figure \ref{figure:124Sn b}(b), the form factors up to  $q \approx 0.7 \;\mathrm{fm}^{-1}$ can be reproduced considering the contributions of fourth-order. Therefore, it is necessary to employ the DWBA method to analyze the contributions of $\left\langle r^{4}\right\rangle_{c}$ of the charge density on the form factors for heavy nuclei.

\subsection{$\left\langle r^{2}\right\rangle_{c}$ and $\left\langle r^{4}\right\rangle_{c}$ extracted from experimental $|F_{C}(q)|^2$ at small $ q $}

As the developments of electron scattering experiments in unstable nuclei, new Coulomb form factors $|F_{C}(q)|^2$ will be gradually measured. At present, due to the experimental conditions are limited, the new experimental data are mainly concentrated at the small momentum transfer $q$ region \cite{PhysRevLett.118.262501}. Therefore, we try to extract the second-order moment $\left\langle r^{2}\right\rangle_{c}$ and fourth-order moment $\left\langle r^{4}\right\rangle_{c}$ from Eqs. (\ref{Fc DW}) and (\ref{D2n nihe}) based on the experimental data at the small $q$ region.

It can be seen from Sec. \ref{c3.2} that when the expansion Eq. (\ref{Fc DW}) takes into account the contributions of the $\left\langle r^{2}\right\rangle_{c}$ and $\left\langle r^{4}\right\rangle_{c}$, the theoretical calculations are sufficient to give the $|F_{C}^{DW}(q)|^2$ in the $q$ region $q < 0.7\; \mathrm{fm}^{-1}$. Therefore, in this section, we expand the $F_{C}^{DW}(q)$ of different nuclei to the fourth-order at the small $q$ region with Eq. (\ref{Fc DW}), and give the relationship between the expansion coefficients ($ D_2, D_4 $) and the proton number $ Z $.

We select 21 candidates from light nuclei to medium mass nuclei with accurate experimental charge density distributions. The $ ^{12} $C, $^{16}$O, $^{28}$Si, $^{32}$S, $^{40}$Ar, $^{48}$Ca, $^{50}$Ti, $^{52}$Cr, $^{56}$Fe, $^{62}$Ni, $^{68}$Zn, $^{72}$Ge, $^{88}$Sr, $^{90}$Zr, $^{92}$Mo, $^{104}$Pd, and $^{144}$Sm used the FB experimental densities \cite{DEVRIES1987495}. The $^{24}$Mg and $^{116}$Sn used the SOG experimental densities \cite{DEVRIES1987495}. The $^{138}$Ba used the Three-parameter Gaussian (3pG) experimental densities \cite{DEVRIES1987495}. The $^{142}$Nd used the Three-parameter Fermi (3pF) experimental densities \cite{DEVRIES1987495}. Based on the experimental charge density, the $|F_{C}(q)|^2$ are calculated by the DWBA method. The second and fourth order expansion coefficients ($ D_2, D_4 $) are obtained from the fourth-order polynomial fitting Eqs. (\ref{Fc DW nihe}) and (\ref{D2n nihe}) with every calculated values of $F_{C}^{DW}(q)$ in the $ q $ range of $0.3\sim0.7 \;\mathrm{fm}^{-1}$. During the fitting procedure, the $\left\langle r^{2}\right\rangle_{c}$ and $\left\langle r^{4}\right\rangle_{c}$ are calculated from the accurate experimental charge density distributions for different nuclei. The fitting results are stable and do not change within the size of the $ q $-range, if the fitting values of $ F_C^{DW}(q) $ are located at  $0.3  \;\mathrm{fm}^{-1}<q<0.7 \;\mathrm{fm}^{-1} $ .

\begin{figure}[t]
	\centering
	\includegraphics[width=9.cm,angle=0]{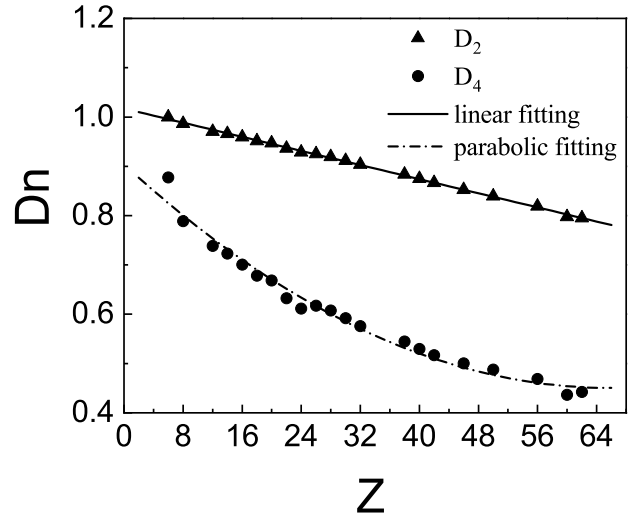}\\
	\caption{  The expansion coefficients $ D_2 $ and $ D_4 $ with proton number $ Z $ for DWBA form factors $|F_{C}^{DW}(q)|^2$. }
	\label{figure:Dn}
\end{figure}

In figure \ref{figure:Dn}, we show the variations of $ D_2 $ and $ D_4 $ with the proton number $ Z $. It can be seen that expansion coefficient $ D_2 $ of DWBA has a noticeable linear relationship, and $ D_4 $ approximately presents a parabolic relation. By fitting the $ D_2 $ and $ D_4 $ we obtain the following relations:

\begin{subequations} \label{DZ}
\begin{align}
	&D_{2}(Z)=1.0172-0.0036 \cdot Z,\label{D2}\\
    &D_{4}(Z)=1.0465 \cdot 10^{-4} \cdot Z^{2}-0.0138 \cdot Z+0.9041.\label{D4}
\end{align}
\end{subequations}

The $ Z $ dependence of the coefficients $ D_2 $ and $ D_4 $ physically reflect the important aspects of the Coulomb distortions. The $ Z $ dependence of $ D_2(Z) $ and $ D_4(Z) $ in Eq. (\ref{DZ}) is consistent with the usual expression for $ q_\mathrm{eff} $. If we expand the second-order $ q_\mathrm{eff}^2 $ in Eq. (\ref{Fc PW}) and keep the first-order term, the $ Z $ dependence of the $ D_2 $ can be obtained. Further expanding the $ q_\mathrm{eff}^4 $ in Eq. (\ref{Fc PW}) and ignoring the high order terms, one can obtain the quadratic dependence on $ Z $ of the $ D_4 $ in Eq. (\ref{D4}). 

It is also meaningful to compare the expansion coefficients of the PWBA form factors and DWBA form factors at the small $ q $ region. For the PWBA form factors at the low $ q $ region, the expansion coefficients $ D_2 $ and $ D_4 $ for each nucleus are both 1, which can be seen in Eq. (\ref{Fc PW}). For the DWBA form factors at the low $ q $ region, the coefficients $ D_2 $ and $ D_4 $ (obtained by Eq. (\ref{DZ})) are close to 1 for light nuclei, and gradually deviate from 1 with the increasing of mass number $ A $. For example, for $^{12}$C, $ D_2 $ = 0.9957 and $ D_4 $ = 0.8252; while for $^{144}$Sm, $ D_2 $ = 0.7952 and $ D_4 $ = 0.4521. This reflects the effects of Coulomb distortion on wave functions of scattered electrons.

With the relationship Eq. (\ref{DZ}) we can extract the $\left\langle r^{2}\right\rangle_{c}$ and $\left\langle r^{4}\right\rangle_{c}$ from the experimental data. For certain nuclei, two arbitrarily experimental data of $|F_{C}(q)|^2$ in the $ q $ region of $0.3\sim0.7 \;\mathrm{fm}^{-1}$ are substituted into Eqs. (\ref{Fc DW}) and (\ref{DZ}), and the $\left\langle r^{2}\right\rangle_{c}$ and $\left\langle r^{4}\right\rangle_{c}$ can be obtained by solving the linear equations. One experimental value is taken near $0.3 \;\mathrm{fm}^{-1} $, and the other is near $0.7 \;\mathrm{fm}^{-1} $. We select 9 nuclei and the results are presented in table \ref{tableIII} to verify the validity of Eq. (\ref{DZ}). The RMF results from the $\mathrm{NL3}^{*}$ parameter set are also shown in table \ref{tableIII} for comparison. 

\begin{table}[t]
	\caption{The second-order moment $\left\langle r^{2}\right\rangle_{c}$ and fourth-order moment $\left\langle r^{4}\right\rangle_{c}$ extracted from the experimental $|F_{C}(q)|^2$ with Eqs. (\ref{Fc DW}) and (\ref{DZ}). The RMF results are obtained with the $\mathrm{NL3}^{*}$ parameter set. The experimental values for Ca, Ni, Zn, Sn isotopes are obtained by analyses of the data in Ref. \cite{DEVRIES1987495}. The experimental $\left\langle r^{2}\right\rangle_{c}$ of Xe is taken from the Ref. \cite{ANGELI201369}. }
	\renewcommand{\arraystretch}{1.1}
	\centering
	\begin{tabular}{lllclclcllclclcl}
		\hline
		\hline
		&        &  & \multicolumn{5}{c}{$\left\langle r^{2}\right\rangle_{c}$} &  &  & \multicolumn{5}{c}{$\left\langle r^{4}\right\rangle_{c}$} &  \\ \cline{4-8} \cline{11-15}
		& Nuclei &  & Extract     &     & $\mathrm{NL3}^{*}$      &     & Expt.     &  &  & Extract    &     & $\mathrm{NL3}^{*}$       &     & Expt.     &  \\ \hline
		& $^{40}$Ca   &  & 12.088      &     & 11.937    &     & 11.902    &  &  & 199.340     &     & 205.224    &     & 199.991   &  \\
		& $^{48}$Ca   &  & 11.916      &     & 11.763    &     & 11.910     &  &  & 197.952    &     & 191.563    &     & 194.714   &  \\
		& $^{58}$Ni   &  & 14.414      &     & 14.028    &     & 14.454    &  &  & 278.805    &     & 267.071    &     & 279.440    &  \\
		& $^{64}$Ni   &  & 15.205      &     & 14.533    &     & 14.603    &  &  & 297.488    &     & 283.710     &     & 283.779   &  \\
		& $^{64}$Zn   &  & 15.462      &     & 15.111    &     & 15.201    &  &  & 299.535    &     & 304.612    &     & 315.308   &  \\
		& $^{70}$Zn   &  & 16.035      &     & 15.421    &     & 15.896    &  &  & 322.242    &     & 317.760     &     & 349.548   &  \\
		& $^{112}$Sn  &  & 21.323      &     & 20.877    &     & 21.030     &  &  & 592.675    &     & 567.005    &     & 586.896   &  \\
		& $^{124}$Sn  &  & 21.878      &     & 21.466    &     & 21.873    &  &  & 626.931    &     & 594.497    &     & 624.196   &  \\
		& $^{132}$Xe  &  & 22.150       &     & 22.520     &     & 22.905    &  &  & 660.512    &     & 647.890     &     &          ------ &  \\ \hline
		\hline
	\end{tabular}
	\label{tableIII}
\end{table}

It can be seen from table \ref{tableIII} that the $\left\langle r^{2}\right\rangle_{c}$ and $\left\langle r^{4}\right\rangle_{c}$ extracted by Eq. (\ref{Fc DW}) are close to the experimental data. The deviations of the extracted $\left\langle r^{2}\right\rangle_{c}^{1/2}$ and $\left\langle r^{4}\right\rangle_{c}^{1/4}$ from the experimental data are both less than 0.08 fm. The accuracy of our method for extracting the $\left\langle r^{2}\right\rangle_{c}^{1/2}$ and $\left\langle r^{4}\right\rangle_{c}^{1/4}$ is within the error range of 2\%. Although the relationships of Eq. (\ref{DZ}) only include the proton number $ Z $, the extracted $\left\langle r^{2}\right\rangle_{c}$ and $\left\langle r^{4}\right\rangle_{c}$ of isotopes can still reflect the effects of neutron number on the charge distributions. The experimental $\left\langle r^{2}\right\rangle_{c}$ and $\left\langle r^{4}\right\rangle_{c}$ in table \ref{tableIII} are obtained by all the experimental data of $|F_{C}(q)|^2$ from small to high $ q $ region. In this paper, we only use two experimental values of $|F_{C}(q)|^2$ at the small $ q $ region to get the reasonable $\left\langle r^{2}\right\rangle_{c}$ and $\left\langle r^{4}\right\rangle_{c}$. Therefore, the reliability of Eq. (\ref{DZ}) can be reflected. 

\begin{figure}[h]
	\centering
	\includegraphics[width=10.cm,angle=0]{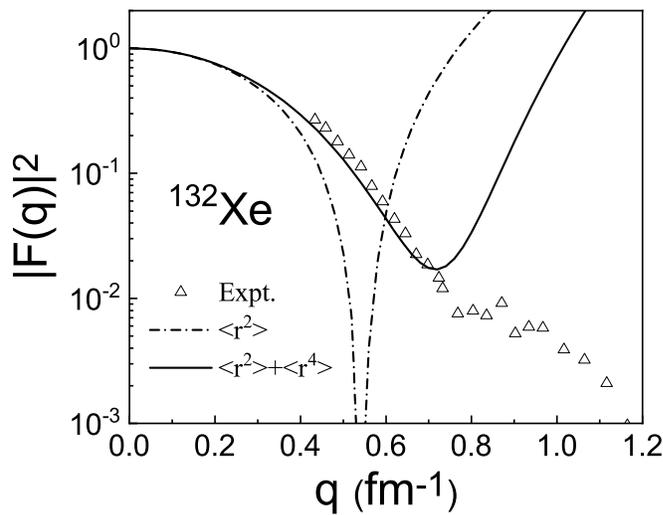}\\
	\caption{ The DWBA form factors $|F_{C}^{DW}(q)|^2$ for $^{132}$Xe are calculated from Eqs. (\ref{Fc DW}) and (\ref{DZ}). The dash-dotted line is extended to the second-order, and the solid line is extended to the fourth-order. The experimental data are taken from Ref. \cite{PhysRevLett.118.262501}. }
	\label{figure:132Xe}
\end{figure}

The $|F_{C}(q)|^2$ of $ ^{132} $Xe is the first electron scattering experiment that has been measured by the SCRIT facility \cite{PhysRevLett.118.262501}. The experimental data of $ ^{132} $Xe are mainly concentrated at the small $ q $ region ($q < 1.5\; \mathrm{fm}^{-1}$). 
By substituting $ Z $ dependence of $ D_{2n} $ of Eq. (\ref{DZ}) and the extracted radial moments in table \ref{tableIII} into Eq. (\ref{Fc DW}), one can obtain he $\left\langle r^{2}\right\rangle_{c}$ and $\left\langle r^{2}\right\rangle_{c}$+$\left\langle r^{4}\right\rangle_{c}$ expansions of the form factors. Figure \ref{figure:132Xe} shows the components of $\left\langle r^{2}\right\rangle_{c}$ and $\left\langle r^{2}\right\rangle_{c}$+$\left\langle r^{4}\right\rangle_{c}$ of the form factors for $ ^{132} $Xe based on the Eq. (\ref{Fc DW}). It can be seen from the figure that only considering the contributions of $\left\langle r^{2}\right\rangle_{c}$, the theoretical values have a noticeable deviation from the experimental data. When adding the contributions of $\left\langle r^{4}\right\rangle_{c}$, the theoretical values are consistent with the experimental data at the small $ q $ region. Therefore, for unstable nuclei, it is feasible to extract the $\left\langle r^{2}\right\rangle_{c}$ and $\left\langle r^{4}\right\rangle_{c}$ from the limited data at the small $ q $ region with Eqs. (\ref{Fc DW}) and (\ref{DZ}). 

\section{Summary and conclusion}\label{c4}

The fourth-order moment $\left\langle r^{4}\right\rangle_{c}$ is a fundamental property of the nuclei associated with the surface thickness of the nuclear density distributions. The corresponding $\left\langle r^{2}\right\rangle_{c}$ and $\left\langle r^{4}\right\rangle_{c}$ can well determine the diffraction radius and surface thickness, especially for the heavy nuclei. Besides, the nuclear neutron radius can also be extracted from the linear relationship between $\left\langle r^{4}\right\rangle_{c}$ and $ R_n^2 $ in the nuclear mean-field calculations. In previous studies, relations between the radial moments $\left\langle r^{n}\right\rangle_{c}$ and form factors $|F_{C}(q)|^2$ at the small momentum transfer $ q $ region were investigated under the framework of the PWBA method. In this paper, we further extend the studies for radial moments and $|F_{C}(q)|^2$ at the small $ q $ region with the DWBA method.

The studies are divided into three parts. Firstly, the theoretical $\left\langle r^{2}\right\rangle_{c}$ and $\left\langle r^{4}\right\rangle_{c}$ are calculated from the different parameterizations and compared with the experimental data, where the corresponding charge distributions are corrected by the contributions of neutron and spin-orbit densities. The corrections of neutron and spin-orbit densities on radial moments are insensitive to parameterizations. Secondly, we present the $|F_{C}(q)|^2$ from the PWBA method and the DWBA method, respectively. One can see that the results of the DWBA method are consistent with the experimental data for heavy nuclei. By expanding the DWBA form factor $F_{C}^{DW}(q)$ into $ q^4 $, one can see the experimental data can be well reproduced by considering the contribution of the $\left\langle r^{4}\right\rangle_{c}$ at the small $ q $ region. Finally, we analyze the second-order and fourth-order expansion coefficients of the $|F_{C}^{DW}(q)|^2$ from light nuclei to medium mass nuclei. Based on the relationship between the expansion coefficients and proton number $ Z $, the $\left\langle r^{2}\right\rangle_{c}$ and $\left\langle r^{4}\right\rangle_{c}$ are extracted from the limited experimental data at the small $ q $ region. The extracted radial moments coincide with experimental data within the allowed error range.

It is challenging to extract $\left\langle r^{4}\right\rangle_{c}$ of exotic nuclei from the elastic electron scattering experiments directly, because $|F_{C}(q)|^2$ can only be measured at the small $ q $ region for exotic nuclei at present. The method proposed in this paper is important for extracting $\left\langle r^{2}\right\rangle_{c}$ and $\left\langle r^{4}\right\rangle_{c}$ based on limited scattering cross sections. The results can also offer useful guides for the coming experiments, which are helpful to interpret the experimental data.

\section*{Acknowledgements}  

The authors are grateful to Toshimi Suda for valuable discussions and careful reading of the manuscript. This work was supported by the National Natural Science Foundation of China (Grants No. 11505292, No. 11775133, No. 11822503, No. 11975167, and No. 12035011), by the Shandong Provincial Natural Science Foundation, China (Grant No. ZR2020MA096), by the Fundamental Research Funds for the Central Universities (Grant No. 20CX05013A, No. 22120210138), and by the Graduate Innovative Research Funds of China University of Petroleum (East China) (Grant No. YCX2020104).

\section*{References}
\bibliography{babo}

\end{document}